\pdfoutput=1
\documentclass[aps,prl,reprint,groupedaddress]{revtex4-1}
\usepackage{graphicx,color}
\usepackage{amsmath, amssymb}
\usepackage{longtable}
\usepackage{natbib}

\begin{document}
\title{
Dynamical Jahn--Teller effect and antiferromagnetism in Cs$_3$C$_{60}$
}
\author{Naoya Iwahara}
\affiliation{Division of Quantum and Physical Chemistry, 
Katholieke Universiteit Leuven, 
Celestijnenlaan 200F, B-3001 Leuven, Belgium}
\author{Liviu F. Chibotaru}
\affiliation{Division of Quantum and Physical Chemistry, 
Katholieke Universiteit Leuven, 
Celestijnenlaan 200F, B-3001 Leuven, Belgium}
\date{\today}

\begin{abstract}
The dynamical Jahn--Teller effect on fullerene sites in insulating Cs$_3$C$_{60}$ is investigated fully ab initio. The vibronic excitations of rotational type are at $\ge$65 cm$^{-1}$ while the net kinetic contribution to the Jahn--Teller stabilization energy constitutes ca 90 meV. This means that no localization of distortions by intermolecular interactions is possible in these fullerides, therefore, free rotations of deformations take place independently on each C$_{60}$. The latter destroy the orbital ordering and establish a conventional exchange interaction between $S=1/2$ on fullerene sites. The corresponding exchange model is derived and predicts the N\'{e}el temperature for A15 Cs$_3$C$_{60}$ close to experiment.
\end{abstract}


\maketitle

%
%
The alkali-doped fullerides A$_3$C$_{60}$ attracted much attention in the past as prominent examples of purely organic high-$T_c$ superconductors \cite{Gunnarsson1997a}. 
Recently a critical temperature of 38 K has been observed in the superconducting Cs$_3$C$_{60}$ \cite{Ganin2008a}, reviving the interest for these fullerides due to their closeness to the Mott-Hubbard metal-insulator transition \cite{Takabayashi2009a,Ganin2010a, Jeglic2009a,Ihara2010a,Kawasaki2013a}. 
It was found that applying an external pressure ($P$) these materials can be brought from insulator to superconductor \cite{Ganin2008a,Takabayashi2009a,Ganin2010a}. 
Such transformation was explained by the increase of the width $w$ of the partly occupied threefold degenerate $t_{1u}$ band under pressure, 
and the concomitant reduction of the ratio $U/w$ ($U$ is the intrafullerene electron repulsion parameter), 
which causes a Mott-Hubbard transition above some critical pressure. 
A thorough investigation of $T_c$ in a wide range of applied pressures revealed its nonmonotonic dependence on pressure (or interfullerene distance) in the metallic phase \cite{Ganin2008a}. 
The maximum of $T_c(P)$ testifies about non-BCS behaviour of Cs$_3$C$_{60}$ close to Mott-Hubbard transition. 
The nonmonotonic behaviour of superconductivity has been qualitatively reproduced by dynamical mean-field calculations \cite{Capone2002a,Han2003a,Capone2009a}, which explained the decline of superconductivity close to Mott-Hubbard transition by the suppression of metallic (and superconducting) fraction of electronic density induced by strong electron correlation \cite{Han2003a}. 
Furthermore, the NMR studies of these compounds  \cite{Jeglic2009a,Ihara2010a,Kawasaki2013a} have evidenced that the maximum of $T_c(P)$ corresponds to the onset of the effects of strong electron correlation on superconductivity. 
Indeed, it was found that the nuclear spin lattice relaxation time $T_1$ for Cs$_3$C$_{60}$ starts to deviate from a BCS-like dependence when the pressure is decreased from the value corresponding to the maximum of $T_c(P)$ towards the Mott-Hubbard transition \cite{Ihara2010a}. At the same time, in the domain of higher pressure the dependence of $T_1$ vs interfullerene distance merges with the smooth curve found on the other A$_3$C$_{60}$ compounds \cite{Maniwa1994a}.    

The similarity of the $T_c(P)$ dependence for Cs$_3$C$_{60}$ with the dome structure of $T_c$ on the $T$-$n$ phase diagram for cuprates has tempted some authors to suppose a close mechanism of superconductivity, based on strong electron correlation, in both these materials. In the case of cuprates and iron pnictides there is a strong evidence that superconducting pairing arises from magnetic fluctuations. However, in the case of fullerides such pairing mechanism is expected to be much less efficient because the exchange interaction in the latter amounts to few meV, i.e., is two orders of magnitude smaller than in cuprates. This points to the electron-phonon coupling as the main contribution to the pairing interaction in Cs$_3$C$_{60}$, 
as was already found for less correlated fullerides K$_3$C$_{60}$ and Rb$_3$C$_{60}$ \cite{Gunnarsson1997a}. The electron-phonon coupling in the LUMO band of fullerides comes mainly from the local vibrations of total-symmetric $a_g$ and Jahn--Teller (JT) $h_g$ type. 
The latter give 4-5 times larger contribution to the stabilization of an electron on fullerene site \cite{Gunnarsson1995a,Iwahara2010a}, the same for the relative contribution to the superconducting pairing \cite{Gunnarsson1997a}. More importantly, it was found that the contribution of JT phonons to the superconducting pairing is suppressed by strong electron correlations at much lesser extent than the contribution from the total symmetric vibrations \cite{Han2003a}, making them a prime source of superconducting pairing in Cs$_3$C$_{60}$. 

The actual role of Jahn--Teller effect (JTE) in fullerides gave rise to controversial opinions. 
It seems to be firmly established nowadays that 
cubic A$_{\text{n}}$C$_{60}$, A=K,Rb,Cs, n=1-6, 
show no static JT distortions in experiments probing their structure. 
Among the recent confirmations, the structural data from the synchrotron x-ray diffraction at 10 K shows 
no JT distortions of C$_{60}^{3-}$ ions in Cs$_3$C$_{60}$ fullerides \cite{Ganin2008a,Takabayashi2009a}. 
The absence of detectable JT distortions in the x-ray data of metallic fullerides was interpreted by many authors as their suppression by band effects. 
At the same time the non-observability of these distortions in insulating A$_4$C$_{60}$ 
is considered to be due to their disordering and partial dynamic delocalization between different minima. 
In expanded cubic fullerides, deeply immersed in a Mott-Hubbard insulating state \cite{Durand2003a}, a dynamical JTE of rotational type on fullerene sites was inferred \cite{Chibotaru2005a}. 
Finally in insulating Cs$_3$C$_{60}$ another type of dynamical JTE, corresponding to fast jumps between different local minima of potential energy surface, was claimed to be the reason for the observed features of the infrared spectra \cite{Klupp2012a}. 
Note that in all these cases the details of JTE have not been known a priori but rather inferred from available experiment \cite{Gunnarsson1995a}. 
In this Letter we present the first fully ab initio treatment of dynamical JTE on C$_{60}^{3-}$ sites in insulating Cs$_3$C$_{60}$ and show its crucial role for the observed magnetism in this fulleride.

%
%

In insulating Cs$_3$C$_{60}$ the electrons from the $t_{1u}$ band become localized at fullerene sites. The $t_{1u}^3$ shell of each trianion C$_{60}^{3-}$ 
splits into three electronic terms, ${^4}S \oplus {^2}P \oplus {^2}D$, at an extent comparable to JT stabilization energy (vide infra). 
Formulated as mixing of the terms, the JT couplings appears only between the $^2P$ and the $^2D$ terms \cite{OBrien1996a}.
Using the complex wave functions of the terms
$\{|^2P,M_P\rangle, |^2D,M_D\rangle; M_P = -1, 0, 1, M_D = -2, -1, 0, 1, 2\}$,
the Hamiltonian matrix is given by 
\begin{eqnarray}
 \hat{H} &=& \hat{H}_0 + \hat{H}_{\rm JT} + \hat{H}_{\rm ee},
\label{Eq:H}
\\
 \hat{H}_0 &=& 
 \sum_{\mu=1}^8 \sum_{m_v=-2}^2 \hslash \omega_\mu 
 \hat{b}_{\mu,m_v}^\dagger \hat{b}_{\mu,m_v}\hat{I},
\label{Eq:H0}
\\
 \hat{H}_{\rm JT} &=&
 \sum_{\mu=1}^8 \frac{\sqrt{3}}{2}\hslash \omega_\mu g_\mu 
 \begin{pmatrix}
  \hat{O}_P & \hat{M}_\mu \\
  \hat{M}_\mu^\dagger & \hat{O}_D\\
 \end{pmatrix},
\label{Eq:HJT}
\\
 \hat{H}_{\rm ee} &=& 2J_{\rm H} \hat{I}_P. 
\label{Eq:Hee}
\end{eqnarray}
Here, $m_v$ denotes the complex basis for $h_g$ vibrations \cite{Auerbach1994a}, 
$\mu$ indexes the $\mu$th $h_g$ vibrational mode,
$\omega_\mu$ is the corresponding frequency,
$\hat{b}_{\mu,m_v}^\dagger$ ($\hat{b}_{\mu,m_v}$) is the creation (annihilation) operator of 
the vibration $\mu h_g m_v$, 
$\hat{I}_\Gamma$ is the projection operator onto the term $\Gamma=P,D$ and 
$\hat{I} = \hat{I}_{P}+\hat{I}_D$, 
$g_\mu$ is the dimensionless orbital vibronic coupling constant, 
$\hat{O}_P$ and $\hat{O}_D$ are the $3 \times 3$ and the $5 \times 5$ zero matrices,
$\hat{M}_\mu$ is defined by ($q_{m_v} \equiv q_{\mu m_v}$)
\begin{eqnarray}
 \hat{M}_\mu &=&
 \begin{pmatrix}
  \sqrt{2}q_{ -1} & \sqrt{3}q_0 &  \sqrt{3}q_{1} & \sqrt{2}q_{2} & 0 \\
  -2q_{-2} & -q_{ -1} & 0 &  q_{1} & 2q_{2} \\
  0 & -\sqrt{2} q_{-2} & -\sqrt{3}q_{-1} & -\sqrt{3} q_{0} & -\sqrt{2}q_{1} \\
 \end{pmatrix},
\label{Eq:M}
\end{eqnarray}
%
%
%
with $q_{\mu,m_v} = 
[\hat{b}_{\mu,m_v}^\dagger + (-1)^{m_v} \hat{b}_{\mu,-m_v}]/\sqrt{2}$,
and $J_{\rm H}$ is the Hund's rule coupling constant.
In Eq. (\ref{Eq:H}) the zero of energy corresponds to the energy of $^2D$ term plus the zero-point vibrational 
energy. 

The JT Hamiltonian (\ref{Eq:H}) commutes with the vibronic angular momentum $\hat{\mathbf{J}}^2$ and
its projection $\hat{J}_z$ \cite{Bersuker1989a}, and with the `parity' operator
\begin{eqnarray}
 \hat{P} &=& \left(\hat{I}_P - \hat{I}_D \right)
 \exp\left(i\pi \sum_{\mu=1}^8 \sum_{m_v=-2}^2 \hat{b}_{\mu,m_v}^\dagger \hat{b}_{\mu,m_v}\right) ,
\label{Eq:P}
\end{eqnarray}
which commute also between themselves.
Then each vibronic state will be characterized the quantum numbers $J (=0,1,2,\cdots)$, 
$M_z (=-J,-J+1,\cdots,J)$ and $P (=\pm 1)$.

As $g_\mu$ in Eq. (\ref{Eq:HJT}), we use the values 
calculated by density-functional theory (DFT) with hybrid B3LYP functional 
and those derived from the photoelectron spectrum (PES) of C$_{60}^-$
(set (2) and (3) in Table S1 of Supplemental Material, respectively\cite{SM}) \cite{Iwahara2010a,comment2}.
Given the good comparison of calculated and measured $g_{\mu}$'s for C$_{60}^-$
($g_\mu$'s derived from several methods are compared in Table V of Ref. \onlinecite{Iwahara2010a})
, we applied the same approach to calculate orbital vibronic coupling constants on fullerene sites in Cs$_3$C$_{60}$ \cite{IwaharaToBePub}. 
To this end a fragment including one C$_{60}^{3-}$ with nearest-neighbour Cs$^+$ ions was considered in the Madelung field of the rest of the crystal. 
The results (set (1) in Table S1) prove to be remarkably close to the case of one isolated C$_{60}^-$ (set (2) in Table S1). This means that the interaction with the lattice and the fullerene charging are insignificant for orbital vibronic constants.
We applied the same DFT approach to evaluate the Hund's coupling constant 
for C$_{60}^{3-}$ and obtained
$J_{\rm H}$ = 44 meV 
which is close to the suggested value of about 50 meV \cite{Martin1993a}.
Then, with experimental frequencies for the eight $h_g$ modes \cite{Gunnarsson1997a}, all terms of the Hamiltonian (\ref{Eq:H}) are completely defined.

The JT Hamiltonian (\ref{Eq:H}) was diagonalized numerically using the Lanczos method as in the case of C$_{60}^-$ \cite{Gunnarsson1995a,Iwahara2010a}. However now the problem is more demanding because of a stronger JTE in C$_{60}^{3-}$ and the need to mix all eight electronic wave functions of the terms $^2P$ and $^2D$. The vibrational wave functions for each of them included linear combinations of products of uncoupled oscillators after 40 $\{\mu h_g m_v \}$ vibrations with total excitation not exceeding seven vibrational quanta.

%
%
\begin{table}[floatfix]
\caption{
Contributions to the ground state vibronic energy of C$_{60}^{3-}$ (meV).
\label{Table:contribution}}
\begin{ruledtabular}
\begin{tabular}{cccccc}
 Set & Total & $\langle \hat{H}_{\rm ee} \rangle$ & JT stabilization & Static & Dynamic \\
\hline
 (2) & -196.2 & 41.0 & -237.2 & -150.6 & -86.6 \\
 (3) & -223.3 & 40.9 & -264.2 & -173.1 & -91.1
\end{tabular}
\end{ruledtabular}
\end{table}
%

%
%
The ground vibronic state corresponds to $J=1$ and $P=1$, i.e., is threefold degenerate after vibronic momentum. Table \ref{Table:contribution} shows that the total stabilization energy with respect to the $^2D$ term (the ground doublet state in the absence of vibronic coupling) is ca 210 meV. This cannot be attributed solely to JT stabilization because JTE in this system implies mixing of $^2D$ and $^2P$ terms leading to the rise of electronic energy, Eq. (\ref{Eq:Hee}), by ca 40 meV. Extracting this contribution from the total energy we obtain the JT stabilization energy of ca 250 meV. This consists of a stabilization energy due to equilibrium static JT distortions (when the kinetic energy of nuclei is neglected \cite{Bersuker1989a}) and due to dynamic delocalization of JT distortions in the three-dimensional trough of the ground state potential energy surface. The former contribution is three times the static JT stabilization energy for a singly charged fullerene \cite{Auerbach1994a} ($E_{\text{JT}}$ in Table S1) and is ca 160 meV. Extracting this value from the total JT stabilization energy we obtain that ca 90 meV corresponds to the dynamic contribution, which is {\em more than half} of the static JT stabilization energy.  

%
%
\begin{table}[floatfix]
\caption{Energies of low-lying vibronic levels in C$_{60}^{3-}$ (cm$^{-1}$) 
calculated for two sets of vibronic parameters.
\label{Table:level}}
\begin{ruledtabular}
\begin{tabular}{ccccc}
 $(J,P)$ & \multicolumn{2}{c}{Set (2)} & \multicolumn{2}{c}{Set (3)} \\
         & exact & effective & exact & effective \\
\hline
$(1,+1)$ &     0 &     0 &     0 &    0 \\
$(2,-1)$ &  65.7 &  65.8 &  63.6 &  63.9 \\
$(3,-1)$ & 254.1 & 453.3 & 251.1 & 427.8 \\
$(4,+1)$ & 283.0 & 393.9 & 273.8 & 369.1 \\
\end{tabular}\end{ruledtabular}
\end{table}

The low-lying vibronic states are characterized by consecutive increase of $J$ (Table \ref{Table:level}). This quantum number corresponds to three-dimensional rotations of JT deformations in the trough of the lowest potential energy surface and its increase with the energy of low-lying levels is generally expected \cite{Bersuker1989a}. However the spacing between these levels differs drastically from the predictions of simplified vibronic models of C$_{60}^{3-}$ \cite{Auerbach1994a,OBrien1996a}. To get more insight into this problem we derive an effective one-mode $t^3 \times h$ JT Hamiltonian (retaining the bielectronic term, Eq. (\ref{Eq:Hee})) that reproduces the static JT stabilization (Table \ref{Table:contribution}) and the energy of the first excited vibronic level. For set 2 (3) we obtain the effective vibronic coupling constant $g_{\text{eff}}$= 1.07 (1.15) and the effective frequency for $h_g$ vibrations $\omega_{\text{eff}}$= 707 (704) cm$^{-1}$. From the obtained $g_{\text{eff}}$ we conclude that the JT effect in C$_{60}^{3-}$ is of intermediate coupling strength. The dynamic contribution to JT stabilization differs by only 11 meV from the exact result. In the strong coupling limit this contribution is (3/2)$\hbar \omega_{\text{eff}}$ \cite{Auerbach1994a,OBrien1996a}, which is ca 50\% higher. The excited states obtained with the effective one-mode JT Hamiltonian do not simulate well the low-lying vibronic spectrum, even give wrong order of levels (Table \ref{Table:level}). This contradicts the general belief that low-lying states of a multimode vibronic problem can be described satisfactorily by an effective one-mode Hamiltonian \cite{OBrien1996a,Chancey1997a}.

The ground vibrational level of non-JT term $^4S$ has the energy $E^{(3/2)}_0 =-3J_{\rm H} = -132$ meV, i.e., lies higher than the ground vibronic state of the spin doublet ($E^{(1/2)}_0$) by 64.2 and 91.3 meV for sets (2) and (3) in Table \ref{Table:contribution}, respectively. While the last value is close to the estimate for the spin gap in this material (0.1 eV) \cite{Jeglic2009a} we note that the activation energy for the spin quartet state should also include the difference of the entropy of vibrational levels for spin quartet ($\ln{Z_{\text{vib}}^{(3/2)}}$) and vibronic levels for spin doublet states ($\ln{Z_{\text{vib}}^{(1/2)}}$):
\begin{equation}
\Delta E^{(3/2)}= E^{(3/2)}_0 -E^{(1/2)}_0 -kT (\ln{Z_{\text{vib}}^{(3/2)}} - \ln{Z_{\text{vib}}^{(1/2)})} .
\label{spin_gap}
\end{equation}
Dynamical JTE in the spin doublet state leads to a denser spectrum of vibronic levels as compared to vibrational spectrum of the spin quartet resulting in $Z_{\text{vib}}^{(1/2)} > Z_{\text{vib}}^{(3/2)}$ for any temperature. Then the entropic term in (\ref{spin_gap}) will increase the activation energy for $T>0$, which thus can rise significantly with temperature. This can explain why no contribution of $S=3/2$ was seen for NMR relaxation times in Cs$_3$C$_{60}$ at room temperature \cite{Ihara2010a}. 

The average value of $E^{(3/2)}_0 -E^{(1/2)}_0$ (ca 80 meV) is comparable to the dynamic contribution to JT stabilization energy (last column in Table \ref{Table:contribution}). This means that if the JT deformations become localized by low-symmetric surrounding of C$_{60}^{3-}$ or random strains, the vibronic levels will approach the vibrational spectrum, so that the dynamic contribution to JT stabilization will be quenched. As a result the spin gap $E^{(3/2)}_0 -E^{(1/2)}_0$ will be strongly reduced, the same for the entropic term in Eq. (\ref{spin_gap}) and the entire activation energy $\Delta E^{(3/2)}$. 
One should note, however, that the low-symmetric environment can itself lead to the splitting of the $t_{1u}$ shell in undistorted fullerenes by several tens of meV \cite{Potocnik2012a} giving and independent contribution to the spin gap, that can thus survive in low-symmetry fullerides.

\begin{figure}
\begin{center}
\includegraphics[bb=0 0 1552 920, height=3cm]{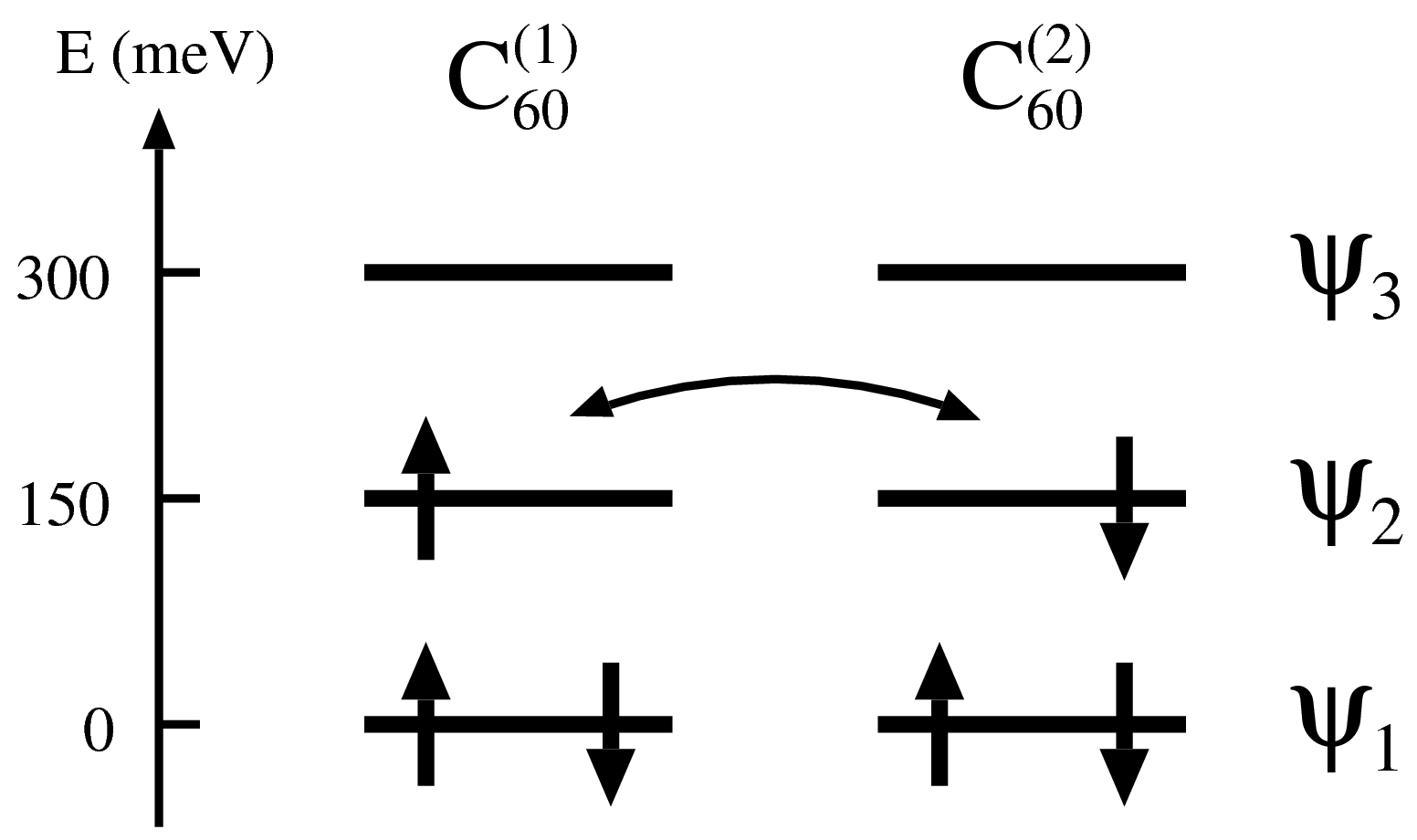}
\end{center}
\caption{
Basic mechanism of a superexchange interaction between fullerene sites in insulating Cs$_3$C$_{60}$.
}
\label{Fig:superexchange}
\end{figure}

%
%
\begin{table*}[htb]
\begin{center}
\caption{
Calculated exchange parameters (meV) and N\'{e}el temperatures and Weiss temperatures (K) for two
sets of transfer integrals (meV), from the present work and from Nomura 
{\it et al}. \cite{Nomura2012a}.
}
\label{Table:Neel}
\begin{ruledtabular}
\begin{tabular}{*{12}{c}}
& $t^{\rm AB}_{xx}$ & $t^{\rm AB}_{yx}$ & $t^{\rm AB}_{zx}$ & $t^{\rm AA}_{xx}$ & $t^{\rm AA}_{yy}$ & $t^{\rm AA}_{zz}$ 
& $w$ & $J_{\rm ex}^{\rm AB}$ & $J_{\rm ex}^{\rm AA}$ & $T_{\rm N}$ & 
{$\Theta$} 
\\
\hline
Present  & -20.1 &  32.5 &   5.6 &   4.8 & -17.6 & -6.7  & 512 & 2.12 & 0.17 & 46 & {-52}\\
Nomura   & -20.6 &  32.9 &   5.3 &   7.3 &  -7.9 & -18.0 & 535 & 2.18 & 0.20 & 48 & {-54}\\
\end{tabular} 
\end{ruledtabular}
\end{center}
\end{table*}

In cubic environment the vibrational $h_g$ modes of C$_{60}$ split into $t_g$ and $e_g$ cubic modes leading to the warping of the adiabatic potential energy surface (APES). The energy difference between the resulting maxima and minima of the lowest APES can be expressed via the frequencies of the corresponding modes ($\omega_{tg}$ and $\omega_{eg}$) as $2[(\omega_{eg}^2-\omega_{tg}^2)/(\omega_{eg}^2+\omega_{tg}^2)]E_{\rm JT}$ \cite{OBrien1996a}. 
Table S2 in Supplemental Material \cite{SM} gives the frequencies of $t_g$ and $e_g$ modes obtained from fragment DFT calculations of Cs$_3$C$_{60}$ \cite{IwaharaToBePub}. 
One can see that the splitting of $h_g$ vibrations in this fulleride does not exceed few wavenumbers. Passing from frequencies in Table S2 to single effective $t_g$ and $e_g$ modes \cite{OBrien1996a,Chancey1997a}, we estimate with the above formula the warping amplitude as 3.1 cm$^{-1}$, which proves to be too small to hinder the free rotations of JT deformations on C$_{60}^{3-}$ (cf. the spacing of low-lying vibronic levels in Table \ref{Table:level}). 
On the other hand the second order JT effect in C$_{60}^{n-}$
\cite{Dunn1995a, Sindi2007a, Lakin2012a, Chancey1997a}
is expected to result in similar amplitude of warping of
the lowest APES, as quantum chemistry calculations show \cite{Koga1992a,Ramanantoanina2012a}, and will not hinder the rotations of JT deformations either \cite{comment3}. 
Finally, the magnetic interaction between fullerenes (of the order of N\'{e}el temperature, $T_{\rm N}$= 46 K \cite{Ganin2008a,Takabayashi2009a,Ganin2010a, Jeglic2009a,Ihara2010a,Kawasaki2013a}) and the intermolecular phonons (with frequencies amounting to several tens of wavenumbers \cite{Gunnarsson1997a}) are comparable in energy with the lowest vibronic excitation (65 cm$^{-1}$) but are still much smaller than the dynamical contribution to JT stabilization energy. We conclude, therefore, that each fullerene in Cs$_3$C$_{60}$ is characterized by {\em unquenched rotations of JT deformations} in the trough of the lowest APES.

Accordingly, the exchange interaction between fullerene sites corresponds to averaging after nuclear coordinates over all points in the three-dimensional trough of each C$_{60}^{3-}$. The way this averaging is done and the resulting exchange mechanism depends on the structure of vibronic state. Overlap integrals of the ground vibronic states with spin doublet configurations $\psi_1^{n_1} \psi_2^{n_2} \psi_3^{n_3}$, $n_1 +n_2 +n_3 =3$, of three adiabatic orbitals $\psi_i$ (the ones which diagonalize the static JT problem for a given JT distortion) show that the ground configuration $\psi_1^{2} \psi_2^{1}$ (see Fig. \ref{Fig:superexchange}) enters with the weight 0.86 and 0.87 for set (2) and (3), respectively. 
Such a high weight of a single adiabatic configuration is rather surprising given that C$_{60}^{3-}$ is far from the strong coupling limit as was mentioned above. 
The main contribution to the exchange interaction between such adiabatic configurations of two fullerene ions comes from superexchange interaction between half filled adiabatic orbitals $\psi_2$ (Fig. \ref{Fig:superexchange}). 
It can be shown \cite{IwaharaToBePub} that in this case the exchange interaction between the lowest states of fullerene sites is of conventional Heisenberg form for spins 
$S=1/2$, with exchange parameters $J^{ij}$ obtained by averaging their values at fixed JT distortions, $J^{ij}(\alpha_i ,\beta_i ,\gamma_i ,\alpha_j ,\beta_j ,\gamma_j )$, over the Euler angles parametrising the angular part of nuclear JT distortions of the two fullerene sites (JT coordinates in the troughs). 
In the case of Anderson's superexchange, which is likely to dominate in fullerides 
\cite{directexchange}, 
the dependence of $J^{ij}$ on the angular coordinates of the troughs enters via the electron transfer parameters between adiabatic orbitals of two fullerene sites \cite{Chibotaru2007a}. Averaging these expressions we obtain the exchange parameters for nearest neighbor (AB) and next-nearest neighbor (AA) C$_{60}^{3-}$ sites corresponding to the main exchange mechanism in Fig. \ref{Fig:superexchange} (see the Supplemental Material):
\begin{eqnarray}
J_1^\text{AB}=\frac{4}{3U^{\text{AB}}} [(t_{xx}^{\text{AB}})^2 +(t_{yx}^{\text{AB}})^2 +(t_{zx}^{\text{AB}})^2 ], \nonumber\\
J_1^\text{AA}=\frac{4}{9} U^{\text{AA}} [(t_{xx}^{\text{AA}})^2 +(t_{yy}^{\text{AA}})^2 +(t_{zz}^{\text{AA}})^2 ],
\label{exchange_main}
\end{eqnarray}
where A and B denote two fullerene sublattices merohedrally rotated with respect
to each other in A15 Cs$_3$C$_{60}$ \cite{Ganin2008a, Takabayashi2009a}; 
$t_{\alpha\beta}$ are electron transfer parameters between orbitals of type $\alpha$ and $\beta$ of two fullerene sites, defined with respect to corresponding tetragonal axes of the crystal; $U=U_{\parallel}-(10/3)J_{\text{H}}+(4/3)E_{\text{JT}}+V_{\text{eh}}$ is the averaged electron promotion energy between two C$_{60}^{3-}$. Next in importance is the exchange interaction coming from virtual electron transfers involving empty and doubly occupied adiabatic orbitals ($\psi_2 \rightarrow \psi_3$, $\psi_1 \rightarrow \psi_2$, $\psi_1 \rightarrow \psi_3$ within adiabatic ground state configurations in Fig. \ref{Fig:superexchange}), which after averaging over JT angular coordinates in the troughs give the following contributions (see the Supplemental Material):
\begin{eqnarray}
J_2^\text{AB}= J_1^\text{AB} \frac{4E_{\text{JT}}-16J_{\text{H}}}{3U^{\text{AB}}}, \nonumber\\
J_2^\text{AA}= J_1^\text{AA} \frac{4E_{\text{JT}}-16J_{\text{H}}}{3U^{\text{AA}}},
\label{exchange_secondary}
\end{eqnarray}
that are ferromagnetic ($J_2 <0$) contrary to contributions (\ref{exchange_main}). The transfer parameters have been evaluated by DFT as matrix elements of Kohn-Sham Hamiltonian between Wannier orbitals constructing  from $t_{1u}$ band orbitals \cite{Nomura2012a} (second row in Table \ref{Table:Neel}). An independent fit of the dispersion of $t_{1u}$ band \cite{IwaharaToBePub} (Fig. S1 in Supplemental Material \cite{SM}) gives an alternative set of transfer parameters (first row in Table \ref{Table:Neel}). 
The electron repulsion within the same $t_{1u}$ orbitals of C$_{60}$ has been calculated by DFT taking into account the RPA screening in insulating Cs$_3$C$_{60}$, $U_{\parallel}$= 1.14 eV, the same for averaged electron-hole attraction for nearest neighbor fullerene sites, 
$V_{\text{eh}}^{\text{AB}}$= -0.34 eV \cite{Nomura2012a}. 
For next-nearest neighbour fullerenes we took the value $V_{\text{eh}}^{\text{AA}}=V_{\text{eh}}^{\text{AB}}R_{\text{AB}}/R_{\text{AA}}$= -0.29 eV, where $R_{\text{AB}}$ and $R_{\text{AA}}$ are the corresponding interfullerene distances. With these values we obtain $U^{\text{AB}}$= 0.72 eV and $U^{\text{AA}}$= 0.77 eV. 
Then with Eqs. (\ref{exchange_main}) and (\ref{exchange_secondary}) we obtain for $J^\text{AB}_\text{ex}=J_1^\text{AB}+J_2^\text{AB}$ and $J^\text{AA}_\text{ex}=J_1^\text{AA}+J_2^\text{AA}$ the values given in Table \ref{Table:Neel}. 
Having in mind that each fullerene has eight nearest and six next-nearest neighbors, 
we calculate the N\'{e}el temperature, $T_{\rm N} =S(S+1) (8J^\text{AB}_\text{ex}-6J^\text{AA}_\text{ex})/3k$, 
and find close agreement with experiment for both sets of transfer parameters (Table \ref{Table:Neel}). 
Similar calculations of Weiss temperature, $\Theta = -S(S+1)(8J_{\rm ex}^{\rm AB} + 6J_{\rm ex}^{\rm AA})/3k$,
give, however, lower absolute values than the experimental one, -68 K (Table \ref{Table:Neel}).

This can be understood by the closeness of metal-insulator transition in Cs$_3$C$_{60}$ at ambient pressure,
which enhances charge transfer fluctuations between fullerene sites,
thus increasing $J_{\rm ex}^{\rm AB}$, $J_{\rm ex}^{\rm AA}$, and $|\Theta|$.
The same reason explains why $\mu_{\rm eff}$ \cite{Takabayashi2009a} extracted from spin susceptibility measurements (1.32)
is smaller than the value corresponding to $S=1/2$ (1.73).
On the other hand the increase of $T_N$ by these fluctuations will be compensated by its decrease due to 
quantum spin fluctuations in the antiferromagnetic phase, thus approaching the mean-field $T_N$ calculated above
to the experimental value.

These calculations give additional evidence for independent rotational dynamics on fullerene sites in insulating Cs$_3$C$_{60}$, since the isotropic Heisenberg model is a fingerprint for such vibronic phase. 
On the contrary, when the vibronic dynamics is quenched, like in low-symmetry insulating (NH$_3$)K$_3$C$_{60}$ \cite{Margadonna2001a}, the half filled orbitals ($\psi_2$ in Fig. \ref{Fig:superexchange}) will be orthogonal for some nearest neighbour fullerenes leading to ferromagnetic exchange interaction similar to $J_2^\text{AB}$ in Eq. (\ref{exchange_secondary}). 
Accordingly, the antiferromagnetic interaction with other fullerene neighbors becomes non-frustrated, resulting in elevated N\'{e}el temperature for this fcc lattice, $T_{\rm N}$= 28 K \cite{Margadonna2001a}. 
For comparison, $T_{\rm N}$ is only 2.2 K in the cubic fcc fulleride Cs$_3$C$_{60}$ \cite{Ganin2010a}, which is explained by isotropic Heisenberg exchange interaction arising in this fulleride due to unquenched dynamical JTE on C$_{60}^{3-}$ sites and, therefore, unavoidable frustration of antiferromagnetic exchange interactions in this lattice.      

N. I. would like to acknowledge the financial support from Flemish Science Foundation (FWO)
and the GOA grant from KU Leuven.


%

\pagebreak
\begin{widetext}
\begin{center}
\textbf{
Supplemental Material\\
 for\\
``Dynamical Jahn--Teller effect and antiferromagnetism in Cs$_3$C$_{60}$''
}
\end{center}
\end{widetext}
\setcounter{equation}{0}
\setcounter{figure}{0}
\setcounter{table}{0}
\makeatletter
\renewcommand{\theequation}{S\arabic{equation}}
\renewcommand{\thefigure}{S\arabic{figure}}
\renewcommand{\thetable}{S\arabic{table}}
\renewcommand{\bibnumfmt}[1]{[S#1]}
\renewcommand{\citenumfont}[1]{S#1}




\begin{table}[tb]
\caption{Dimensionless vibronic coupling constants $g_{\mu}$ and
JT stabilization energies from individual modes $E_{{\rm JT},\mu}$ (meV) for C$_{60}^-$.
Set (1) is obtained from fragment DFT calculations of Cs$_3$C$_{60}$ and
sets (2) and (3) are derived from DFT calculations
and PES of C$_{60}^-$, respectively \cite{SMIwahara2010a}.}
\label{Table:g}
\begin{ruledtabular}
\begin{tabular}{ccccccc}
      & \multicolumn{2}{c}{Set (1)} & \multicolumn{2}{c}{Set (2)} & \multicolumn{2}{c}{Set (3)} \\
$\mu$ & $g_\mu$ &  $E_{\rm JT,\mu}$ & $g_\mu$ &  $E_{\rm JT,\mu}$ & $g_\mu$ & $E_{\rm JT,\mu}$\\
\hline
1 & 0.413 & 2.9 & 0.436 & 3.2 & 0.490 & 4.1 \\
2 & 0.484 & 6.3 & 0.498 & 6.7 & 0.515 & 7.2 \\
3 & 0.437 & 8.4 & 0.418 & 7.7 & 0.455 & 9.1 \\
4 & 0.258 & 3.2 & 0.259 & 3.2 & 0.300 & 4.3 \\
5 & 0.210 & 3.0 & 0.211 & 3.0 & 0.280 & 5.3 \\
6 & 0.137 & 1.5 & 0.126 & 1.2 & 0.235 & 4.3 \\
7 & 0.383 &13.0 & 0.398 &14.0 & 0.435 &16.8 \\
8 & 0.332 &10.8 & 0.338 &11.2 & 0.260 & 6.6 \\
$E_{\rm JT}$ & & 49.1 & &50.2 &       &57.7 \\
\end{tabular}
\end{ruledtabular}
\end{table}

Frequencies of C$_{60}$ and A15 Cs$_3$C$_{60}$ are shown in Table \ref{Table:freq}.
For the estimation of the splitting of $h_g$ vibrational modes by the 
cubic environment of each fullerene in Cs$_3$C$_{60}$, 
the frequencies of the $e_g$ and the $t_g$ modes are used.
The frequencies of an isolated neutral C$_{60}$ are given 
(i) to prove that our calculations of frequencies are reliable
and (ii) to see how much the charging effect changes the frequencies.
The obtained frequencies of C$_{60}$ are close to the experimental ones \cite{SMGunnarsson1997a}
and to DFT values obtained by other authors \cite{SMGunnarsson1997a, SMChoi2000a}.

\begin{table}[tb]
\caption{Frequencies of $h_g$ vibrational modes (cm$^{-1}$) of isolated C$_{60}$ and in A15 Cs$_3$C$_{60}$
derived from DFT calculations.}
\label{Table:freq}
\begin{ruledtabular}
\begin{tabular}{ccc}
C$_{60}$ & \multicolumn{2}{c}{A15 Cs$_3$C$_{60}$} \\
 $h_g$ & $t_g$ & $e_g$ \\
\hline
 264.8 &  271.4 &  271.7 \\
 435.0 &  427.1 &  431.7 \\
 721.1 &  696.6 &  699.4 \\
 784.5 &  786.2 &  786.1 \\
1123.0 & 1117.5 & 1117.9 \\
1265.2 & 1249.0 & 1250.1 \\
1442.1 & 1448.0 & 1449.2 \\
1608.3 & 1598.2 & 1599.7 \\
\end{tabular}
\end{ruledtabular}
\end{table}

%
%
\begin{figure}
\begin{center}
\includegraphics[viewport=0 0 3728 6152, height=8cm, angle=-90]{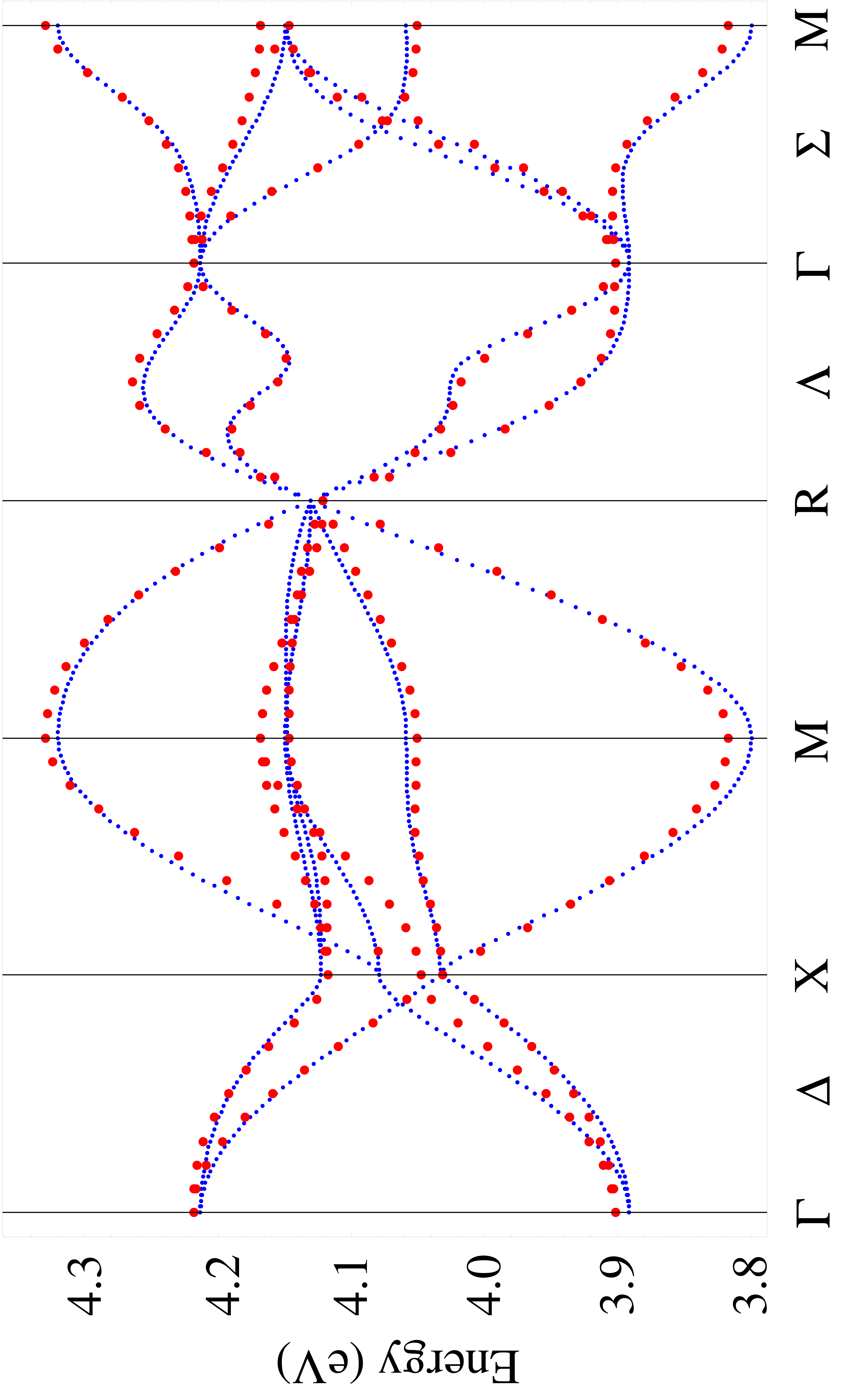}
\end{center}
\caption{
(color online)
Band structure of A15 Cs$_3$C$_{60}$ obtained by DFT calculation (red circles) and 
the effective model with parameters in Table IV (blue dots).
}
\label{Fig:Band}
\end{figure}

%
%

\noindent
{\bf Derivation of Eqs. (8) and (9)\cite{SMChibotaru2007a}}

Suppose the Jahn--Teller deformation of each fullerene is fixed at first. 
The Hamiltonian for two sites ($i,j$) is 
\begin{eqnarray}
 \hat{H} &=& \hat{H}_t + \hat{H}_{\rm JT} + \hat{H}_{bi} + \hat{H}_{\rm eh}, 
\\
 \hat{H}_t &=& \sum_{\lambda \lambda'}\sum_\sigma t_{\lambda \lambda'} 
 \left(\hat{c}_{i\lambda \sigma}^\dagger \hat{c}_{j\lambda'\sigma} 
     + \hat{c}_{j\lambda' \sigma}^\dagger \hat{c}_{i\lambda\sigma}
 \right),
\\
 \hat{H}_{\rm JT} &=& \sum_{m = i,j} 3E_{\rm JT}\sum_\sigma \left(\hat{n}_{m 3\sigma} - \hat{n}_{m 1\sigma}\right) + 3E_{\rm JT}, \\
 \hat{H}_{bi} &=& \sum_{m = i,j} \frac{1}{2} \sum_{\lambda = 1}^3 \sum_\sigma
 \left[
   U_\parallel \hat{n}_{m\lambda\sigma} \hat{n}_{m\lambda,-\sigma} 
 \right.
\nonumber\\
 &+&  
  U_\perp \sum_{\lambda' (\ne \lambda)} \sum_{\sigma'} \hat{n}_{m\lambda \sigma} \hat{n}_{m\lambda \sigma'}
 -J_{\rm H} \sum_{\lambda'(\ne \lambda)} 
 \left(
  \hat{n}_{m\lambda \sigma} \hat{n}_{m\lambda' \sigma}  
 \right.
\nonumber\\
 &-& 
 \hat{c}_{m\lambda \sigma}^\dagger \hat{c}_{m\lambda'\sigma} 
 \hat{c}_{m\lambda,-\sigma}^\dagger \hat{c}_{m\lambda',-\sigma}
\nonumber\\
 &-& 
 \left.
 \left.
 -\hat{c}_{m\lambda \sigma}^\dagger \hat{c}_{m\lambda'\sigma} 
  \hat{c}_{m\lambda',-\sigma}^\dagger \hat{c}_{m\lambda,-\sigma}
 \right) 
 \right],
\\
\hat{H}_{\rm eh} &=& -V_{\rm eh} \sum_{\lambda \sigma} \hat{n}_{i\lambda \sigma} \sum_{\lambda' \sigma'} \hat{n}_{j\lambda' \sigma'}, 
\end{eqnarray}
where $m (= i,j)$ is the index for site, 
$\lambda, \lambda' (= 1,2,3)$ the indices for adiabatic orbital (see Fig. 1 in the text), 
$\sigma (= \uparrow, \downarrow)$ the spin projection, 
$\hat{c}_{m\lambda \sigma}^\dagger$ and $\hat{c}_{m\lambda \sigma}$ the creation and annihilation operators, respectively, 
$\hat{n}_{m\lambda \sigma}$ the number operator, 
$E_{\rm JT}$ the Jahn--Teller stabilization energy for C$_{60}^-$, 
$t_{\lambda \lambda'}$ the transfer parameter, 
$U_\parallel$ the intraorbital onsite Coulomb repulsion, 
$U_\perp (=U_\parallel - 2J_{\rm H})$ the interorbital onsite Coulomb repulsion, 
$J_{\rm H}$ is the Hund's coupling, 
and $V_{\rm eh} (<0)$ is the intersite electron-hole attraction.

We regard $\hat{H}_{bi} + \hat{H}_{\rm JT} + \hat{H}_{\rm eh}$ as unperturbed Hamiltonian and 
$\hat{H}_t$ as perturbation.
The unperturbed Hamiltonian is divided into two terms:
first term $\hat{H}_0$ depends only on the sum of number operators, 
the other $\hat{H}'$ consists of remaining terms.
\begin{widetext}
\begin{eqnarray}
 \hat{H}_0 &=& \sum_{m=i,j} \left[\frac{U}{2}\sum_{\lambda \sigma}
               \left(\hat{n}_{m\lambda\sigma} \hat{n}_{m\lambda,-\sigma} 
               + \sum_{\lambda' (\ne \lambda)} \sum_{\sigma'} \hat{n}_{m\lambda \sigma} \hat{n}_{m\lambda'\sigma'}
               \right) + E_{\rm JT}\right] + \hat{H}_{\rm eh}, 
\label{Eq:H0SM}
\\
 \hat{H}'  &=& \sum_{m=i,j}\left\{ E_{\rm JT}\sum_\sigma \left(\hat{n}_{m 3\sigma} - \hat{n}_{m 1\sigma}\right)
           + \frac{A}{2}
               \sum_{\lambda \sigma} \hat{n}_{m\lambda \sigma} \hat{n}_{m\lambda,-\sigma}
           - \frac{1}{2} \sum_{\lambda \sigma} 
           \left[
            B\sum_{\lambda'(\ne \lambda)} \sum_{\sigma'} \hat{n}_{m\lambda\sigma} \hat{n}_{m\lambda'\sigma'}
           \right.
           \right.
\nonumber\\
           &+& 
            \left.
            \left.
               J_{\rm H} \sum_{\lambda'(\ne \lambda)} 
               \left(
                 \hat{n}_{m\lambda \sigma} \hat{n}_{m\lambda' \sigma}  
                - \hat{c}_{m\lambda \sigma}^\dagger \hat{c}_{m\lambda'\sigma} \hat{c}_{m\lambda,-\sigma}^\dagger \hat{c}_{m\lambda',-\sigma}
                - \hat{c}_{m\lambda \sigma}^\dagger \hat{c}_{m\lambda'\sigma} \hat{c}_{m\lambda',-\sigma}^\dagger \hat{c}_{m\lambda,-\sigma}
                \right) 
           \right]
           \right\},
\label{Eq:H'}
\end{eqnarray}
\end{widetext}
where $U$ is an promotion energy 
\begin{eqnarray}
 U &=& U_\parallel - \frac{10}{3}J_{\rm H} + \frac{4}{3}E_{\rm JT} + V_{\rm eh},
\label{Eq:U}
\end{eqnarray}
$A$ is 
\begin{eqnarray}
 A &=& \frac{10}{3}J_{\rm H} - \frac{4}{3}E_{\rm JT},
\end{eqnarray}
and $B$ is 
\begin{eqnarray}
 B &=& -\frac{4}{3}J_{\rm H}+\frac{4}{3}E_{\rm JT}.
\end{eqnarray}

Performing a unitary transformation, we remove $\hat{H}_t$ in the first order.
\begin{eqnarray}
 \hat{H}_{\rm eff} &=& \hat{H}_0 + \hat{H}' + \frac{1}{U}\hat{H}_t \hat{H}_t - 
 \frac{1}{2U^2}\left[\left[\hat{H}', \hat{H}_t\right],\hat{H}_t\right].
\nonumber\\
\label{Eq:Heff}
\end{eqnarray}
Since spin-orbit coupling is not included, orbital and spin degrees of freedom can be separated.
The interaction between site $i$ and site $j$ is 
\begin{eqnarray}
 \hat{H}^{ij} &=& \hat{K}^{ij} + \hat{J}^{ij}\left(\mathbf{S}_i\cdot \mathbf{S}_j + \frac{1}{2}\right).
\label{Eq:Hij}
\end{eqnarray}
Here, $\hat{K}^{ij}$ and $\hat{J}^{ij}$ are operators which act on orbital space, and $\mathbf{S}$ is the spin operator.

To obtain the superexchange Hamiltonian, $\hat{H}_{\rm eff}$ is averaged by the ground vibronic states of the fullerenes.
First, $\hat{H}_{\rm eff}$ is averaged by electronic part of the vibronic state:
\begin{eqnarray}
 \hat{H}_{\rm ex} &=& K^{ij} + \left(J^{ij}_1 + J^{ij}_2\right) \left(\mathbf{S}_i\cdot \mathbf{S}_j + \frac{1}{2}\right),
\end{eqnarray}
where $K^{ij}$ is 
\begin{eqnarray}
 K^{ij} &=& -2\frac{\left(t_{13}^{ij}\right)^2+\left(t_{31}^{ij}\right)^2}{U}\left(1 - \frac{2E_{\rm JT} + J_{\rm H}}{3U}\right)
\nonumber\\
        &-&\frac{\left(t_{12}^{ij}\right)^2+\left(t_{21}^{ij}\right)^2}{U}\left(1 - \frac{E_{\rm JT} - J_{\rm H}}{3U}\right)
\nonumber\\
        &-& \frac{\left(t_{23}^{ij}\right)^2+\left(t_{32}^{ij}\right)^2}{U}\left(1 + \frac{E_{\rm JT} - 7J_{\rm H}}{3U}\right)
\nonumber\\
        &-&2\frac{\left(t_{22}^{ij}\right)^2}{U}\left(1 + \frac{4E_{\rm JT} - 10J_{\rm H}}{3U}\right),
\end{eqnarray}
$J^{ij}_1$ and $J^{ij}_2$ are 
\begin{eqnarray}
 J^{ij}_1 &=& 4\frac{\left(t_{22}^{ij}\right)^2}{U}, 
\label{Eq:Jij1}
\\
 J^{ij}_2 &=& 4\frac{\left(t_{22}^{ij}\right)^2}{U}\frac{4E_{\rm JT}-10J_H}{3U}
\nonumber\\
             &-&2\frac{\left(t_{12}^{ij}\right)^2+\left(t_{21}^{ij}\right)^2+\left(t_{23}^{ij}\right)^2+\left(t_{32}^{ij}\right)^2}{U}\frac{J_H}{U},
\label{Eq:Jij2}
\end{eqnarray}
respectively. 
$K^{ij}$ and $J^{ij}$ depend on the Euler angles ($\alpha, \beta, \gamma$) parametrizing the angular part of the Jahn--Teller deformation. 

To complete the averaging of $\hat{H}_{\rm eff}$ by the ground vibronic states, 
$J^{ij}$'s are averaged over the Euler angles.
The transfer parameters $t_{\lambda \lambda'}$ between nearest neighbour sites (AB) 
and next nearest neighbours (AA) are described by $t_{\alpha \beta}$ ($\alpha, \beta = x,y,z$) as follows:
\begin{widetext}
\begin{eqnarray}
 \begin{pmatrix}
  t^{\rm AB}_{11} & t^{\rm AB}_{12} & t^{\rm AB}_{13} \\
  t^{\rm AB}_{21} & t^{\rm AB}_{22} & t^{\rm AB}_{23} \\
  t^{\rm AB}_{31} & t^{\rm AB}_{32} & t^{\rm AB}_{33} \\
 \end{pmatrix}
 &=&
B_P(\alpha_i)
C_P(\beta_i)
D_P(\gamma_i)
\begin{pmatrix}
 t^{\rm AB}_{xx} & t^{\rm AB}_{zx} & t^{\rm AB}_{yx} \\
 t^{\rm AB}_{yx} & t^{\rm AB}_{xx} & t^{\rm AB}_{zx} \\
 t^{\rm AB}_{zx} & t^{\rm AB}_{yx} & t^{\rm AB}_{xx} 
\end{pmatrix}
D_P^\dagger(\gamma_j)
C_P^\dagger(\beta_j)
B_P^\dagger(\alpha_j),
\\
 \begin{pmatrix}
  t^{\rm AA}_{11} & t^{\rm AA}_{12} & t^{\rm AA}_{13} \\
  t^{\rm AA}_{21} & t^{\rm AA}_{22} & t^{\rm AA}_{23} \\
  t^{\rm AA}_{31} & t^{\rm AA}_{32} & t^{\rm AA}_{33} \\
 \end{pmatrix}
 &=&
B_P(\alpha_i)
C_P(\beta_i)
D_P(\gamma_i)
\begin{pmatrix}
 t^{\rm AA}_{xx} & 0 & 0 \\
 0 & t^{\rm AA}_{yy} & 0 \\
 0 & 0 & t^{\rm AA}_{zz} \\
\end{pmatrix}
D_P^\dagger(\gamma_j)
C_P^\dagger(\beta_j)
B_P^\dagger(\alpha_j),
\end{eqnarray}
\end{widetext}
where
\begin{eqnarray}
 B_P(\alpha) &=& 
 \begin{pmatrix}
  \cos \alpha & \sin \alpha & 0 \\
 -\sin \alpha & \cos \alpha & 0 \\
  0 & 0 & 1\\
 \end{pmatrix},
\\
 C_P(\beta) &=& 
 \begin{pmatrix}
  \cos \beta & 0 & -\sin \beta \\
  0 & 1 & 0\\
  \sin \beta & 0 & \cos \beta \\
 \end{pmatrix},
\\
 D_P(\gamma) &=& 
 \begin{pmatrix}
  \cos \gamma & \sin \gamma & 0 \\
 -\sin \gamma & \cos \gamma & 0 \\
  0 & 0 & 1\\
 \end{pmatrix}.
\end{eqnarray}
The averages of the squared transfer parameters $t_{\lambda \lambda'}^2$ are given by 
\begin{eqnarray}
 \langle \left(t_{\lambda \lambda'}^{\rm AB}\right)^2\rangle 
&=& \frac{1}{4\pi^4}\int_0^\pi d\alpha_i \int_0^{\pi/2} d\beta_i \sin \beta_i \int_0^{2\pi} d\gamma_i
\nonumber\\
 &\times&
 \int_0^\pi d\alpha_j \int_0^{\pi/2} d\beta_j \sin \beta_j \int_0^{2\pi} d\gamma_j \left(t_{\lambda \lambda'}^{\rm AB}\right)^2
\nonumber\\
\\
 &=& \frac{\left(t_{xx}^{\rm AB}\right)^2+\left(t_{yx}^{\rm AB}\right)^2+\left(t_{zx}^{\rm AB}\right)^2}{3},
\label{Eq:tAB}
\\
 \langle \left(t_{\lambda \lambda'}^{\rm AA}\right)^2\rangle 
&=& \frac{1}{4\pi^4}\int_0^\pi d\alpha_i \int_0^{\pi/2} d\beta_i \sin \beta_i \int_0^{2\pi} d\gamma_i
\nonumber\\
 &\times&
 \int_0^\pi d\alpha_j \int_0^{\pi/2} d\beta_j \sin \beta_j \int_0^{2\pi} d\gamma_j \left(t_{\lambda \lambda'}^{\rm AA}\right)^2
\nonumber\\
\\
 &=& \frac{\left(t_{xx}^{\rm AA}\right)^2+\left(t_{yy}^{\rm AA}\right)^2+\left(t_{zz}^{\rm AA}\right)^2}{9}.
\label{Eq:tAA}
\end{eqnarray}
With use of Eq. (\ref{Eq:tAB}), the superexchange amplitude for the nearest neighbour is 
\begin{eqnarray}
 J_{\rm ex}^{\rm AB} &=& J_{1}^{\rm AB} + J_{2}^{\rm AB},
\label{Eq:JexAB}
\\
 J_1^{\rm AB} &=& \frac{4}{3U_{\rm AB}} 
 \left[\left(t_{xx}^{\rm AB}\right)^2+\left(t_{yx}^{\rm AB}\right)^2+\left(t_{zx}^{\rm AB}\right)^2\right],
\label{Eq:J1AB}
\\
 J_2^{\rm AB} &=& J_1^{\rm AB} 
 \frac{4E_{\rm JT} - 16 J_{\rm H}}{3U_{\rm AB}}.
\label{Eq:J2AB}
\end{eqnarray}
Similarly, using Eq. (\ref{Eq:tAA}), the superexchange amplitude for the next nearest neighbour is 
\begin{eqnarray}
 J_{\rm ex}^{\rm AA} &=& J_{1}^{\rm AA} + J_{2}^{\rm AA},
\label{Eq:JexAA}
\\
 J_1^{\rm AA} &=& \frac{4}{9U_{\rm AA}} 
 \left[\left(t_{xx}^{\rm AA}\right)^2+\left(t_{yy}^{\rm AA}\right)^2+\left(t_{zz}^{\rm AA}\right)^2\right],
\label{Eq:J1AA}
\\
 J_2^{\rm AA} &=& J_1^{\rm AA} 
 \frac{4E_{\rm JT} - 16 J_{\rm H}}{3U_{\rm AA}}.
\label{Eq:J2AA}
\end{eqnarray}
Eqs. (\ref{Eq:J1AB}) and (\ref{Eq:J1AA}) are Eqs. (8),
and Eqs. (\ref{Eq:J2AB}) and (\ref{Eq:J2AA}) are Eqs. (9) in the text.



\begin{thebibliography}{34}
\bibitem{directexchange}
The contribution of direct (potential) exchange is expected to be small because of small overlap
density of magnetic orbitals. 
The latter can only be important when the magnetic orbitals at two sites have significant hybridization
with some bridging atoms \cite{Kahn}.
This is not the case in the ionic lattice of Cs$_3$C$_{60}$ where the Mulliken charge of Cs is very close
to $+1$ (0.995) as obtained from our DFT calculations.
\end{thebibliography}

\end{document}